# GBT20, a 20.48 Gbps PAM4 Optical Transmitter Module for Particle Physics Experiments

B. Deng,[a] L. Zhang,[b,c] C.-P. Chao,[d] S.-W. Chen,[d] E. Cruda,[b] D. Gong,[b] S. Hou,[e] G. Huang,[c] X. Huang,[b] C.-Y. Li,[d] C. Liu,[b] T. Liu,[b, 1] E.R. Liu,[f] Q. Sun,[g] X. Sun,[c] G. Wong,[h] and J. Ye[b]

[a] *Hubei Polytechnic University,*
 *Huangshi, Hubei 435003, P.R. China*

[b] *Southern Methodist University,*
 *Dallas, TX 75275, U.S.A.*

[c] *Central China Normal University,*
 *Wuhan, Hubei 430079, P.R. China*

[d] *APAC Opto Electronics Inc.,*
 *Hukow, Hsinchu 303, Taiwan*

[e] *Academia Sinica, Nangng,*
 *Taipei 11529, Taiwan*

[f] *Allen High School,*
 *TX 75002, U.S.A.*

[g] *Fermi National Accelerator Laboratory,*
 *Batavia, IL 60510, U.S.A.*

[h] *Plano East Senior High School,*
 *Plano, TX 75074, U.S.A*

 E-mail: `tliu@smu.edu`

ABSTRACT: We present a pluggable radiation-tolerant 4-level Pulse-Amplitude-Modulation (PAM4) optical transmitter module called GBT20 (Giga-Bit Transmitter at 20 Gbps) for particle-physics experiments. GBT20 has an OSFP or firefly connector to input 16-bit data each at 1.28 Gbps. The GBT20 drives a VCSEL die with an LC lens or a VCSEL TOSA and interfaces an optical fiber with a standard LC connector. The minimum module, including the host connector, occupies 41 mm × 13 mm × 6 mm. At 20.48 Gbps, the minimum Transmitter Dispersion Eye Closure Quaternary (TDECQ) is around 0.7 dB. The power consumption is around 164 mW in the low-power mode. The SEE cross-section is below $7.5 \times 10^{-14}$ cm$^2$. No significant performance degrades after a TID of 5.4 kGy.

---

[1] Corresponding author.

## 1. Introduction

High-speed optical links are commonly used in modern particle-physics experiments to transmit large-volume data from detectors to counting rooms [1-5]. Future particle-physics experiments have a great demand for high-bandwidth and high-reliability optical links [6]. Based on a radiation-tolerant PAM4 (4-level Pulse-Amplitude-Modulation) ASIC GBS20 [7], a transmitter optical module called the Giga-Bit Transmitter at 20 Gbps (GBT20) is designed for optical links in particle-physics experiments.

## 2. ASIC overview

The block diagram of the GBS20 ASIC is shown in figure 1. The GBS20 has 16 input data channels, which are split into the Least Significant Bit (LSB) and Most Significant Bit (MSB) channels. Each input data channel operates at 1.28 Gbps. By employing phase aligners, we ensure that the input data are sampled correctly. The input data are optionally scrambled (XOR'ed) with a $2^7$-1 Pseudo-Random Binary Sequence (PRBS) generated by an internal test-pattern generator before being fed to two serializers. After passing through a five-stage Limiting Amplifier (LA), the output data of the LSB or MSB channel are encoded into a PAM4 signal by a PAM4 Combiner [8]. GBS20 operates at 10.24 or 20.48 Gbps. The PAM4 combiner directly drives a Vertical-Cavity-Surface-Emitting-Laser (VCSEL) diode. A low-jitter Phase-Locked Loop (PLL), which is derived from lpGBT [9], supplies a few clocks of different frequencies for the phase aligners, the encoders, and two serializers. An Inter-Integrated Circuit ($I^2C$) target block is implemented to configure GBS20. The power supply of the entire chip is divided into three parts: digital 1.2 V, analog 1.2 V, and analog 2.5 V. The 2.5 V power domain is used only in the PAM4 Combiner to expand the modulation current. The supply voltage of the Combiner can be reduced to 1.2 V in a low-power mode.

## 3. Module design

The block diagram of the GBT20 module is shown in figure 2. We adopt two kinds of gold fingers (a 60-pin OSFP and a 48-pin Firefly) as the electrical signal interface. A GBS20 ASIC die is wire bonded to the module Printed Circuit Board (PCB) and is AC coupled to a VCSEL diode. The DC bias circuit of the VCSEL diode is implemented on the module. The GBT20 module uses either a die (Part No. APA45001010001 from II-VI Laser Enterprise) or a Transmitter-Optical-Subassembly (TOSA) (Part No. TTL-1F67-627 from Truelight) of a VCSEL diode. The VCSEL die is directly wire-bonded to the PCB, and a lens with an LC ferrule receptacle (Part No. OT002 from Orangetek Corporation, we will call it an LC lens in this paper) is mounted above the VCSEL die for optical coupling. The TOSA is assembled to the PCB with a flexible cable. The LC lens or TOSA provides the interface with a standard LC connector. A custom latch attaches the LC connector to the LC lens or TOSA.

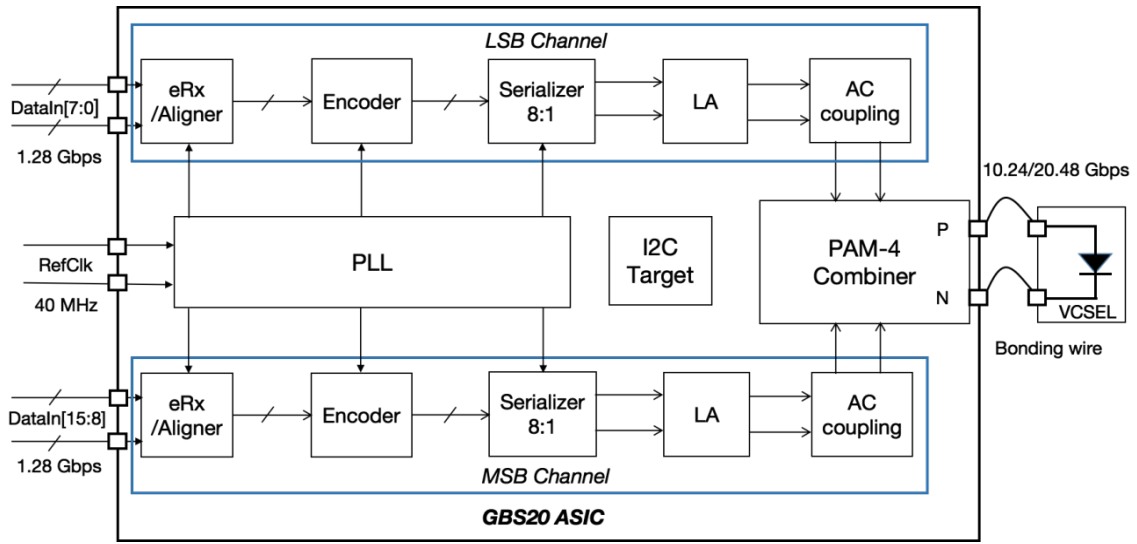

**Figure 1**. Block diagram of the VCSEL transmitter ASIC GBS20.

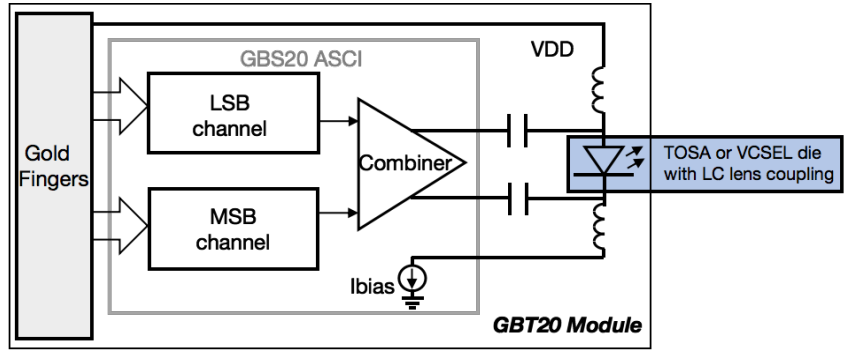

**Figure 2**. Block diagram of the GBT20 module.

The photographs of GBT20 modules with different VCSEL diodes (a die or TOSA) and electrical connectors (OSFP or firefly) are shown in figure 3. In the rest of this paper, we will call a component with a TOSA (die, OSFP connector, or firefly connector) a TOSA (die, OSFP, or firefly) module. For the die modules, the VCSEL dies, which are underneath the LC lenses, are invisible in figure 3. GBS20 chips are protected with plastic covers on the OSFP modules or black encapsulation on the firefly modules. The TOSA modules are higher and longer than the die modules, and the firefly modules are smaller than the OSFP modules. The height of the OSFP connector limits the height of the OSFP modules. The firefly connector is much lower than the OSFP connector. The height of the firefly modules is restricted by the LC ferrule receptacle. In figure 3, a fiber with an LC connector is plugged into the custom latch of each module. All the latches are printed from composite-x resin with a 3-D printer. The tensile strength of the latch is 50–75 MPa after UV curing. Custom screws fix the module PCB and the latch to the motherboard. Table 1 lists the size of the GBT20 modules. The module with a firefly connector and a TOSA occupies the smallest space (45 mm × 13 mm × 6 mm), including the host connector mounted on the motherboard.

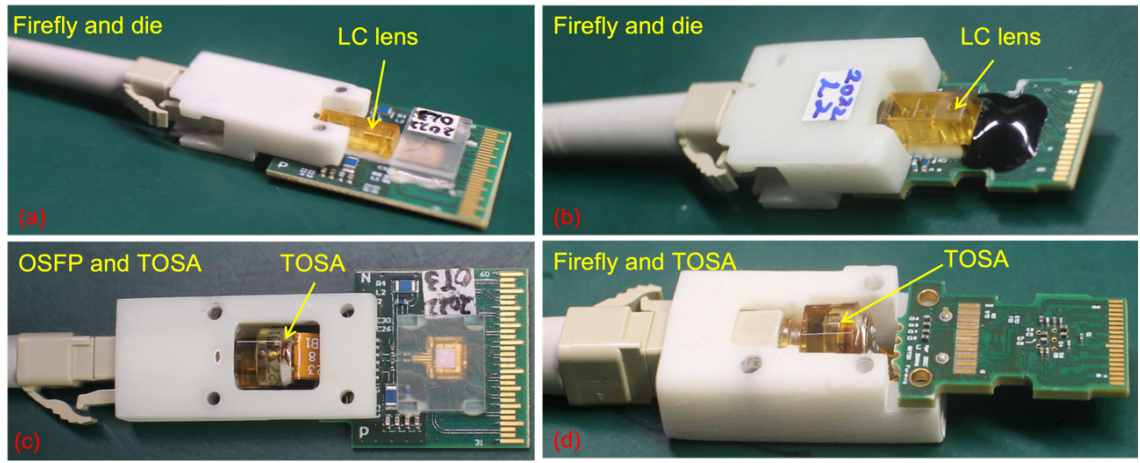

**Figure 3**. Photographs of GBT20 modules with a die and an OSFP connector (a), a die and a firefly connector (b), a TOSA and an OSFP connector (c), a TOSA and a firefly connector (d).

**Table 1.** Size of GBT20 modules (L × W × H, unit: mm).

| Connector or VCSEL type | OSFP | Firefly |
|---|---|---|
| die | 48 × 21.6 × 8.5 | 41 × 13 × 7 |
| TOSA | 52 × 21.6 × 8.5 | 45 × 13 × 6 |

The photographs of an OSFP or firefly module plugged into the motherboard are shown in figure 4. Note that the motherboard is an FPGA Mezzanine Card (FMC). To take advantage of the height room of the OSFP connector, the OSFP modules are plugged into the motherboard with the latch located on the bottom side of the module PCB, compared with the firefly modules.

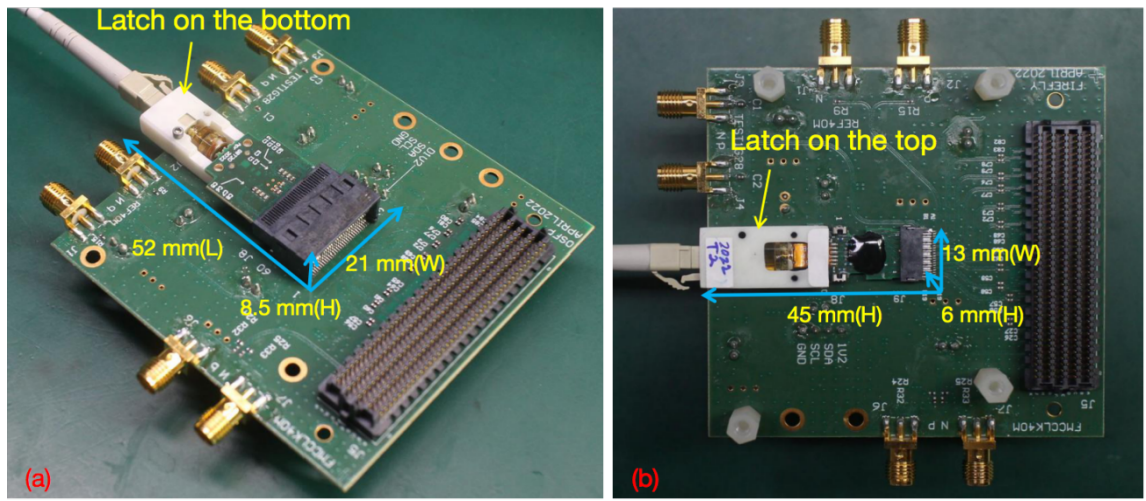

**Figure 4**. Photographs of a module with an OSFP and a TOSA (a) or a firefly and a TOSA (b) plugged into a motherboard.

## 4. Module test

The block diagram and a photograph of the test setup are shown in figure 5. A Cyclone 10 GX FPGA (Part No. DK-DEEV-10CX220-A from Intel) generated a 16-bit input differential PRBS (pattern $2^7-1$) or a repeating pattern of 01010101. The input data went through an FMC connector before being fed to the motherboard with a GBT20 module plugged in. The output of the GBT20 module was connected to either an optical oscilloscope (Model TDS8000B from Tektronix shown in the figure or Model N1000-Series from Keysight) to observe eye diagrams or another FPGA (KC705 from Xilinx) to measure the Bit Error Rate (BER). A clock builder board (Part No. Si5338EVB from Silicon Labs) provided three clocks. The first clock supplied the 40 MHz reference clock of the GBS20 ASIC. The second clock was at 40 MHz for the Cyclone 10 GX FPGA. The other clock was a 40 MHz trigger signal for the oscilloscope or a 160 MHz clock for the FPGA KC705. We used a USB-to-I$^2$C adapter (Part No. ISS-USB from Robot-Electronics) with an I$^2$C level translator (Part No. PCA9306 from Onsemi) or the Cyclone 10 GX FPGA to configure the GBT20 module under test.

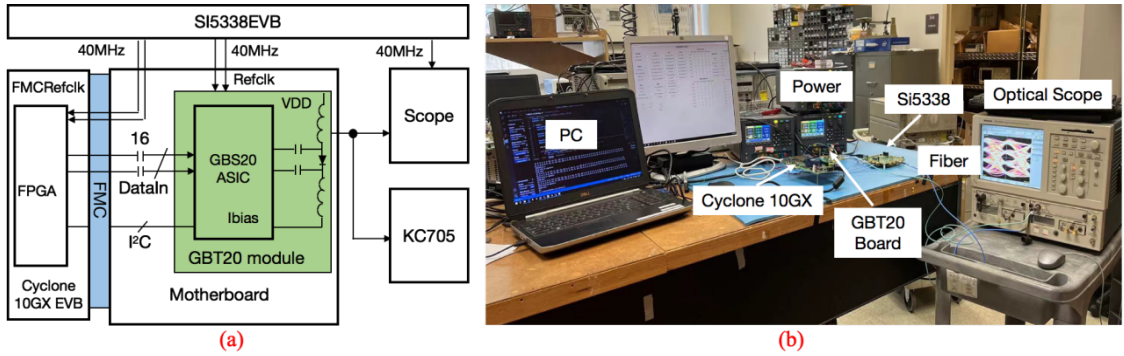

**Figure 5**. Block diagram (a) and photograph (b) of the test setup.

Figure 6 shows 20.48 Gbps PAM4 optical eye diagrams for different types of modules. The outer Optical Modulation Amplitude (OMAouter) is around 1.3 mW for the die modules and 0.8 mW for the TOSA modules, respectively. We notice that all our modules use the same VCSEL diodes. The fact that the OMAouter of the die modules is higher than that of the TOSA modules indicates that the LC lens has a higher coupling efficiency than the TOSA. The outer Extension Ratios (ERs) are all above 6 dB. The Transmitter Dispersion Eye Closure Quaternary (TDECQ) [10] is 1.5 dB for the die modules and 0.7 dB for the TOSA modules, respectively. The reason why the TOSA modules have a better TDECQ than the die modules is still under investigation. The power dissipation is 250 mW in the nominal operational mode and is reduced to 164 mW in the low-power mode.

Due to the limit of the test equipment, the bit error rate was measured only for the Non-Return-to-Zero (NRZ) signal of GBT20. No error was detected for three hours at 10.24 Gbps in the lab environment. The BER is estimated to be $2.7\times10^{-14}$ at a confidence level of 95%.

An irradiation test was conducted at the Irradiation Test Area of Fermi National Accelerator Laboratory (FNAL) in Chicago, USA. The photos of the irradiation test setup are shown in figure 7. About $7.5\times10^{11}$ protons of 400 MeV were delivered within a spill of 10 μs every minute during the test. The fluxes that the chip received were $8.3\times10^{15}$ p/cm$^2$/s when the chip was aligned with the center of the beam and $2.0\times10^{13}$ p/cm$^2$/s when the chip was 67 mm away from the center of the beam, respectively. No Single-Event Effect (SEE) was observed during a fluence of $1.3\times10^{13}$ protons/cm$^2$. The cross-section of the SEE is below $7.5\times10^{-14}$ cm$^2$. No significant performance degradation was observed after a Total Ionizing Dose (TID) of 5.4 kGy.

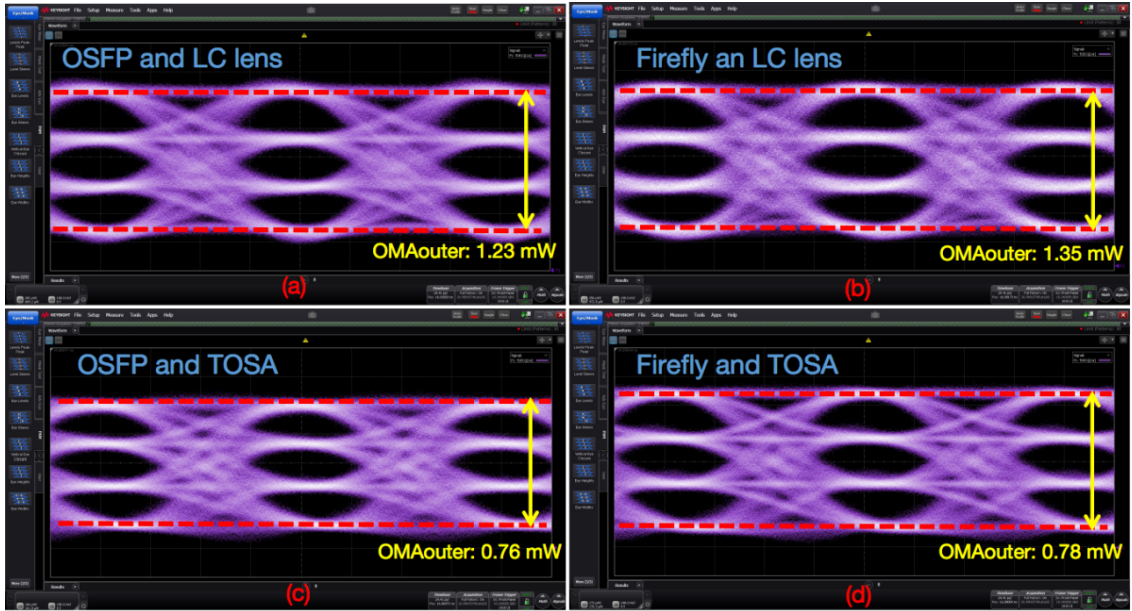

**Figure 6**. PAM4 eye diagrams of modules at 20.48 Gbps with an OSFP connector and a die (a), a firefly connector and a die (b), an OSFP connector and a TOSA (c), and a firefly connector and a TOSA (d).

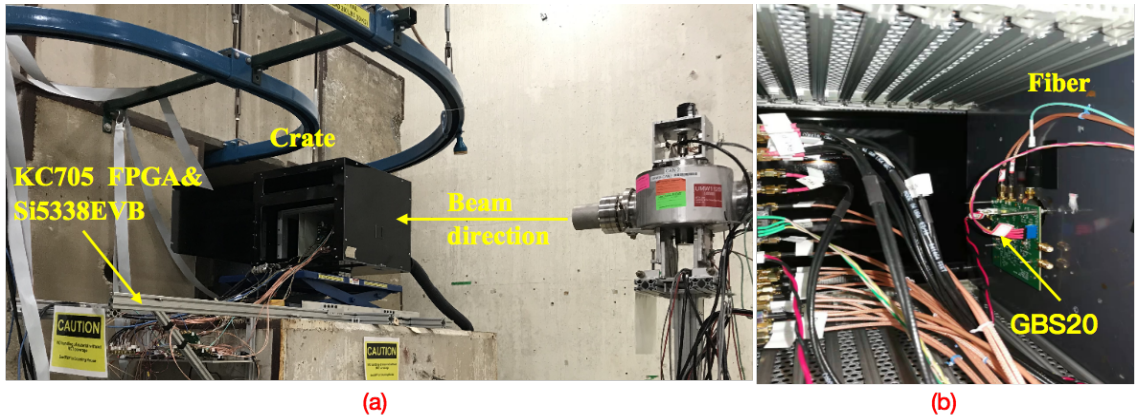

**Figure 7**. Photographs of the irradiation test setup (a) and the test board inserted into the crate (b).

## 5. Summary and outlook

We present the design and test results of GBT20 for particle-physics experiments based on the ASIC GBS20. The GBT20 has an OSFP or firefly connector for 16-bit input data each at 1.28 Gbps. GBT20 drives a VCSEL die with an LC lens or a VCSEL TOSA and interfaces an optical fiber with a standard LC connector. The minimum module, including the host connector, occupies 41 mm × 13 mm × 6 mm. At 20.48 Gbps, the minimum TDECQ is around 0.7 dB. The power consumption is around 164 mW in the low-power mode. The SEE cross-section is below $7.5\times10^{-14}$ cm$^2$. No significant performance degrades after a TID of 5.4 kGy.

## Acknowledgments

The work is supported by SMU's Dedman Dean's Research Council Grant and the National Science Council in Taiwan.